# THERMAL DESIGN OF POWER SEMICONDUCTOR MODULES FOR MOBILE COMMNICATION SYSYTEMS

*Yasuo Osone\**

\*Mechanical Engineering Research Laboratory, Hitachi, Ltd., 832-2 Horiguchi, Hitachinaka, Ibaraki, 312-0034, JAPAN; yasuo.osone.td@hitachi.com

## ABSTRACT

We investigated the thermal performance of a power semiconductor module used in mobile communication systems. The module contained hetero-junction bipolar transistor (HBT) fingers, which were fabricated on the top of a semiconductor substrate. We calculated the thermal resistance between the HBT fingers and the bottom surface of a multi-layer printed circuit board (PCB) using a finite element method (FEM) simulation. We used steady state analysis to evaluate the effect of various design factors on the thermal resistance of the module. We found that the thickness of the semiconductor substrate, the thickness of the multi-layer PCB, the thermal conductivity of the bonding material under the semiconductor substrate, and misalignment of thermal vias between each layer of PCB were the most important factors affecting the thermal resistance of the module.

## 1. INTRODUCTION

A radio frequency (RF) communication circuit is an electrical component used in mobile communication systems. In a cell phone, about half of the phone's electrical power consumption is accounted for by the heat generate by the power amplifier, which amplifies the transmission signal from a transistor, in the phone's RF circuit.

In the thermal design of power amplifiers, the objective is to develop a package structure of the module where the temperature does not exceed an allowable limit. Since power amplifiers have become progressively smaller, their thermal design has been a key factor in developing cell phones.

Other key factors are the need to decrease turn around times in product development and to find a numerical method to optimize the thermal structure of products, which will be of significant use for thermal design work. Modeling know-how is also important I thermal design because automated modeling is difficult to apply to semiconductor devices.

The objective of this paper is to show the thermal design technique of power amplifier modules in call phones using a finite element method (FEM). We evaluated the effect of various design factors on the thermal resistance of the power amplifier module. The thermal design of cell phones is becoming more important because they are approaching the limits of the allowable amount of heat that can be discharged into the environment. In fact, the heat dissipation densities of smaller power amplifiers have increased to 100 W/cm$^2$.

The problem is that, unlike when modeling an automobile, automatic modeling is not easily applied to the numerical modeling of semiconductor modules since their design data - such as mask layout data - are two-dimensional.

Further, the modeling has to handle a huge dynamic range. This range extends from the thickness of a 10 nm film to the 10 cm characteristic length of a printed circuit board. By applying FEM, thermal design, which involves optimizing the packaging structure, predicting the device temperature, and reducing the device temperature so that is does not exceed the allowable limit, is feasible.

By applying an FEM technique, we investigated the thermal performance of power amplifier modules with GaAs hetero-junction bipolar transistors (HBTs). The thermal resistance of HBTs had been investigated using various approaches[1][2][3][4][5]. We believe that numerical methods are most appropriate for developing heat-dissipating structures with low thermal resistance. In this study, we considered not only the layout and the shape of the HBT fingers, and the thickness and thermal conductivity of each component in the power amplifier module but also the misalignment of each component.

The results of the simulation showed that that the thickness of GaAs substrate, the thickness of multi-layer PCBs, and the thermal conductivity of bonding material under the GaAs substrate were the dominant factors affecting the thermal resistance of the power amplifier module. The results also showed that the misalignment of GaAs substrate on the heat spreader did not have a major effect on the thermal resistance of the power amplifier module.

## 2. THERMAL DESIGN OF CELL PHONES

Figure 1 shows a schematic diagram for a cell phone of the flow path of heat between heat sources and ambient air. Heat dissipated by discrete heat sources assembled on a PCB is first transferred to the PCB, and then conducted through the PCB to the front and rear panels by contact heat transfer. The heat in the front and rear panels is conducted to the skin of the cell phone user.

The thermal design of semiconductor modules, such as power amplifiers, baseband large scale integrated (LSI) circuits, application processors, and other devices, usually considers the thermal resistance between the heat source and the PCB. Though the flow path of heat illustrated in fig. 1 and the thermal design model of heat flowing between heat sources and the PCB is not the same, the difference is negligible small because the heat capacity of the panels is much larger than that of the PCB.

Ram et al. stated that the maximum amount of heat dissipated in a handheld computer or cell phone is about 5 W under average conditions[6]. If the total heat dissipation by semiconductors in the cell phone under such average conditions is lower than the maximum amount of heat, the thermal resistance of the front and rear panel is negligible. Since the heat capacity of the panels is much larger than those of other components in the cell phones, the interface between the PCB and the panels can be assumed that the isothermal boundary in the thermal design of the semiconductors in the cell phone.

Figure 2 shows a flow chart of the thermal design process of power semiconductor modules mounted on a PCB when the effect of the case is negligible. The objective of thermal design of a module is to reduce the thermal resistance Rjc between the heat source and the ambient air so that the thermal resistance is lower than the allowable limit. Though both steady state and transient analysis are ideally required for the design, the transient analysis is usually omitted because the transient temperature increase is lower than that of the steady state when the heat dissipation from the heat sources is at a maximum. The steady state analysis will be discussed in the following section.

## 3. ANALYTICAL MODEL OF POWER AMPLIFIER MODULE

Figure 3 shows the package structure of the power amplifier module surveyed in this paper. The thickness of the semiconductor substrate in fig. 3 is about 0.1 mm. Semiconductor devices, such as the HBT fingers, are fabricated on the top surface of the substrate.

The package structure of the power amplifier module in this paper has a plated heat sink (PHS) that is plated under the bottom surface of the semiconductor substrate. The heat dissipated from the HBT fingers diffuses through the semiconductor substrate in fig. 3 horizontally to the PHS, the bonding paste-1, and the heat spreader. The temperature distribution in the flow path of heat is not constant even while it goes through the PHS layer.

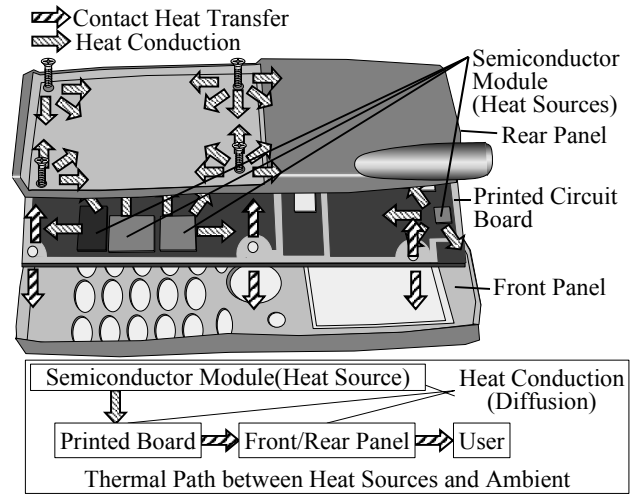

Figure 1 Schematic diagram illustrating flow path of heat between heat sources and ambient in a cell phone

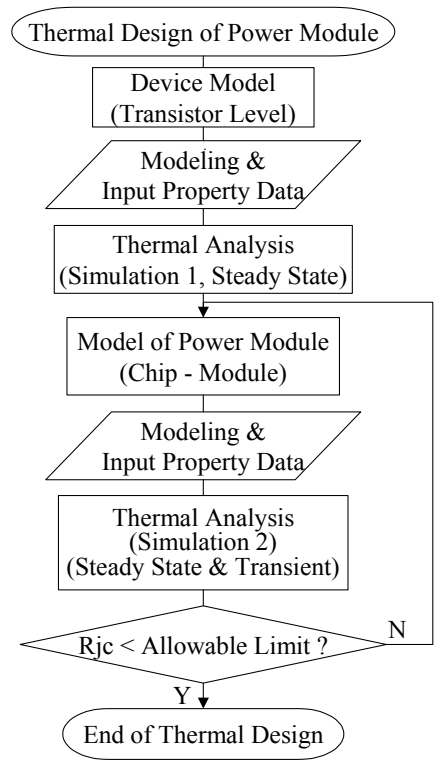

Figure 2 Flow chart illustrating the thermal design of power semiconductor modules on PCB

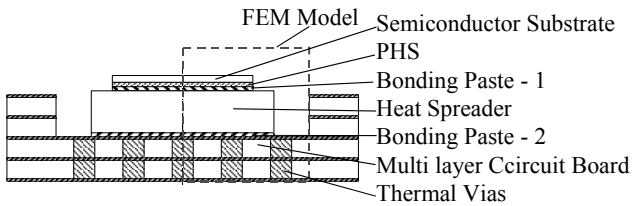

Figure 3 Cross cutting model of a power amplifier module

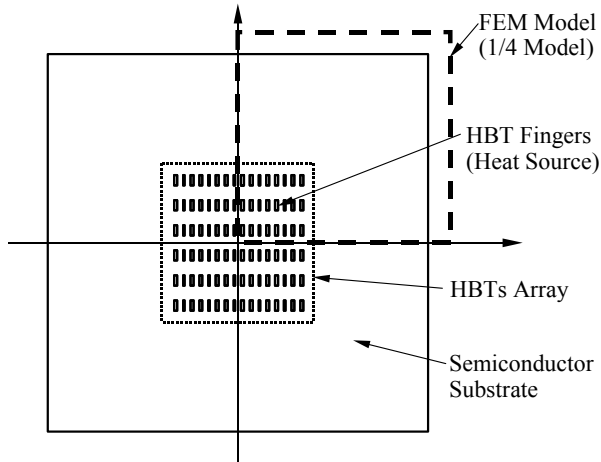

Figure 4 Quarter area model for the FEM analysis

Table 1 Thermophysical properties of materials

| Material | Density [kg/m$^3$] | Specific heat [kJ/(kg-K)] | Thermal conductivity [W/(m-K)] |
|---|---|---|---|
| Insulator (SiO$_2$) | 2190 | 0.860 | 1.5 |
| Semiconductor substrate (GaAs) | 5320 | 0.350 | 46.0 |
| PHS | 19300 | 0.131 | 313.0 |
| Heat spreader (Cu) | 8930 | 0.397 | 395.0 |
| Thermal via | 8400 | 0.290 | 270.0 |
| Base material of multi-layer PCB (ceramic) | 3200 | 0.800 | 2.5 |
| Bonding paste-1 | 4500 | 0.320 | 30.0 |
| Bonding paste-2 | 14520 | 0.200 | 24.0 |

The module package structure is composed of the semiconductor substrate with HBT fingers, a PHS, the bonding paste-1, a heat spreader, bonding paste-2, and a multi-layer PCB. The multi-layer PCB has thermal vias, which are thermally conductive and penetrate the multi-layer PCB. Most of the heat dissipated by the HBT fingers goes through the thermal vias to the external boundary of the module because the multi-layer PCB is primarily made of heat-insulating ceramic material.

Figure 4 shows the HBT fingers layout on the semiconductor substrate. To simplify the analysis, we used an orthogonal coordinate system. The shape of each layer was transformed into a rectangular parallelepiped. We assumed that the finger pitch was constant and the center position of the fingers coincided with the center of the semiconductor substrate so that a quarter area model could be used in the FEM analysis to reduce the size of the simulation model. We also assumed that the isothermal temperature at the bottom surface of the model in fig. 3 was 273 K. Other boundary conditions were adiabatic. The thermophysical properties of the materials used in the analysis are shown in Table 1.

We determined the thermal resistance of the HBT fingers at the point where the temperature difference between the bottom surface and the place in which the temperature was higher than any other place in the model. We calculated the thermal resistance of the HBT fingers, the influence of the thickness of each component in the module, thermophysical properties of the heat spreader, and the alignment of the semiconductor substrate on the multi-layer PCB. In the thermal analysis, we used an HBT model that had 96 HBT fingers (16 fingers x 6 columns) and two layers of multi-layer PCB.

We used a massively parallel processing computer (HITACHI SR-8000 Model G1) for the thermal analysis. The theoretical peak performance of the computer was 230.4 GFLOPS, and the total size of the memories of the system was 128 GB. The node size of the model in the analysis was from 1 to 10 M nodes for the thermal design of the power amplifier module.

## 4. RESULTS OF THERMAL ANALYSIS

**4.1. Thermal resistance of the power amplifier module**
Figure 5 shows the influence of the thickness of the GaAs substrate and the PHS layer on the thermal resistance of the power amplifier module with HBT fingers. As shown in fig. 5.1, we found that the thermal resistance of the power amplifier module could be reduced by 5.3 % compared to that of the typical model shown in fig. 5.1, if the thickness of the GaAs substrate was reduced to 75 % of the typical thickness of the GaAs substrate.

The thermal resistance of the power amplifier module was positively correlated with the thickness of the GaAs substrate, whereas it was negatively correlated with the thickness of the PHS layer. Figure 5.2 shows that the PHS served as a heat spreader and the in-plane heat diffusion in the PHS layer had a significant effect on the reduction of the thermal resistance of the power amplifier module.

**4.2. Impact of thickness of ceramic layer and misalignment on thermal resistance**
Figure 6 shows the impact of the number of ceramic layers in the multi-layer PCB on the thermal resistance of the power amplifier module. In fig. 6, "-100 %" means that the module had no multi-layer PCB and the bottom surface of the heat spreader is isothermal at 273 K. Since the multi-layer PCB shown in fig. 3 has two ceramic layers, "-50 %" means it has one ceramic layer, and "+50 %" means it has three ceramic layers.

We found that the number of ceramic layers in the multi-layer PCB had the greatest impact on the thermal resistance of the power amplifier module. The principal flow path of heat in the multi-layer PCB was through the thermal vias. The heat dissipated in the semiconductor substrate diffused one dimensionally through the thermal vias. Therefore, the thermal resistance of the power amplifier module had a linear dependence on the thickness of the multi-layer PCB, i.e., the number of ceramic layers in the multi-layer PCB.

We found that effect of the misalignment of GaAs substrate on the heat spreader on the thermal resistance was not as significant as the other evaluation items. Since the heat spreader diffuses heat through the bonding paste-1 three dimensionally, and the temperature distribution on the top surface of the bonding paste-2 is almost flat even if the location of the GaAs substrate is misaligned, any misalignment of the GaAs substrate seems to have a small effect on the thermal resistance of the power amplifier module.

The flow path of heat in the multi-layer substrate is one-dimensional. Therefore, the misalignment of the thermal vias should have a major effect on the thermal resistance of the power amplifier module. The degree of the impact of the misalignment of the thermal vias on the thermal resistance depends on the base material of the multi-layer PCB, the production method of the circuit board, and the difference of the thermal conductivity of the base material and the thermal vias.

Since most RF module manufacturers procure PCBs for their modules from PCB venders, they are not able to reduce the misalignment of PCB components by themselves. Therefore, improving the robustness of module thermal design will be an important way to reduce the negative effect of misalignment of each component in the module.

### 4.3. Impact of evaluation items on thermal resistance

Figure 8 shows the impact of each evaluation item on the thermal resistance of the semiconductor module. The most important factor among these evaluation items was the number of the ceramic layers in the multi-layer PCB.

The second most important factor was the thermal conductivity of the bonding paste-1. The thermal conductivity of the bonding paste is important because the impact of the thermophysical properties of the bonding paste can include the influence of the interface thermal resistance between the bonding paste and the bonded material. Even if the thermal conductivity of the bonding material is high enough as a thermally conductive material, the substantive thermal conductivity with the effect of the interface should be considered in the thermal design of the power amplifier module.

We also found that the thermal resistance of the module could be increased to 125 % of that of a typical model.

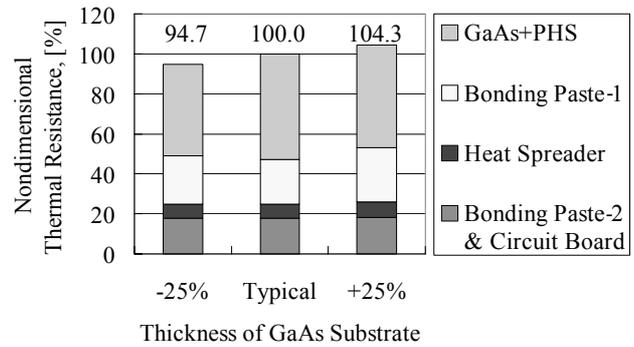

Figure 5.1 Impact of thickness of GaAs substrate

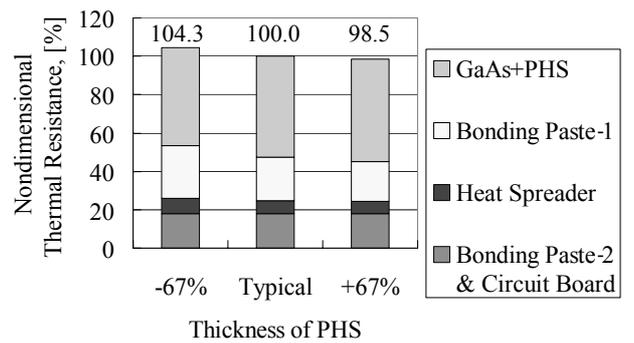

Figure 5.2 Impact of thickness of PHS

Figure 5 Impact of thickness of GaAs substrate, PHS on thermal resistance

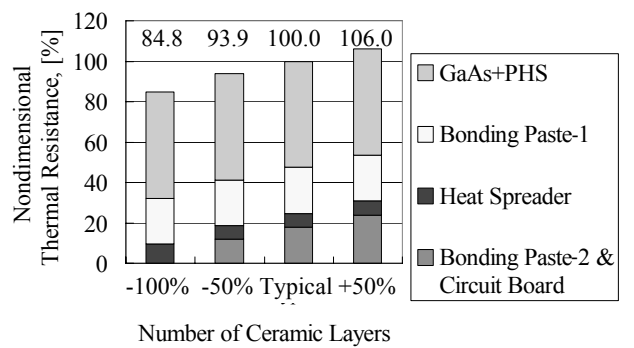

Figure 6 Impact of number of ceramic layer in multi-layer PCB on thermal resistance

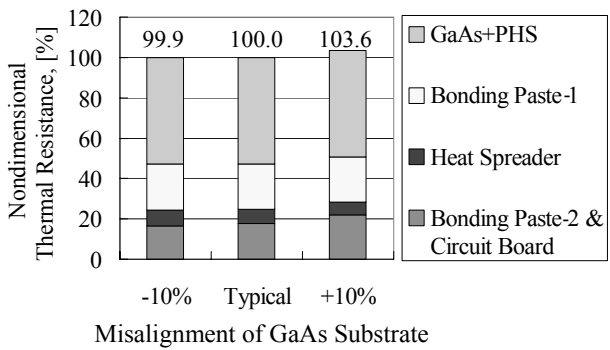

Figure 7 Impact of misalignment of GaAs substrate on thermal resistance

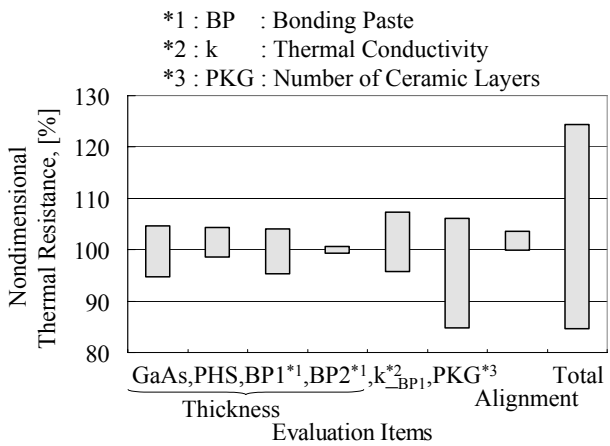

Figure 8 Impact of each evaluation item on thermal resistance

## 5. SUMMARY

We used FEM to investigate the thermal performance of a power amplifier module that had HBT fingers on a GaAs substrate. The effect on the thermal performance of not only the layout and the shape of the HBT fingers and thickness, and thermal conductivity of each component in the module, but also the misalignment of each component was determined.

The results of the FEM simulation showed that that the thickness of GaAs substrate, the thickness of multi-layer PCB, and the thermal conductivity of bonding material under the GaAs substrate were the dominant factors in affecting the thermal resistance of the power amplifier module. The results also showed that the misalignment of GaAs substrate on the heat spreader did not have a major effect on the thermal resistance of the power amplifier module.

The results also showed that the three dimensional shape of thermal vias are important in the thermal design of power amplifier modules because the flow path of heat of the thermal vias in the multi-layer PCB is one-dimensional.